\documentclass{revtex4}    
\usepackage{amssymb,epsf}
\usepackage{latexsym}
\begin{document}
\title{Gravitating Isovector Solitons}
\author{Nematollah \surname{Riazi}}
\address{Physics Department and Biruni Observatory,
\\Shiraz University,
Shiraz 71454, Iran,\\
and\\
IPM, Farmanieh, Tehran, Iran.\\
email: {\rm riazi@physics.susc.ac.ir}}
\author{Hassan \surname{Niad}}
\address{Physics Department and Biruni Observatory,
\\Shiraz University,
Shiraz 71454, Iran. }
%
\begin{abstract}
We formulate the nonlinear isovector model in a curved background,
and calculate the spherically symmetric solutions for weak and
strong coupling regimes. The usual belief that gravity does not
have appreciable effects on the structure of solitons will be
examined, in the framework of the calculated solutions, by
comparing the flat-space and curved-space solutions. It turns out
that in the strong coupling regime, gravity has essential effects
on the solutions. Masses of the self-gravitating solitons are
calculated numerically using the Tolman expression, and its
behavior as a function of the coupling constant of the model is
studied.
\end{abstract}
\maketitle

\section{Introduction\label{sec1}}
There has been considerable interest in the localized solutions of
the Einstein's equations with nonlinear field sources in recent
years (\cite{1}, \cite{2}, \cite{3}, \cite{4}, \cite{5}).
Gravitating non-abelian solitons and black holes with Yang-Mills
fields is investigated in \cite{5b}. Such problems were not
investigated earlier in the history of GR, mainly because of two
reasons: 1. It was widely accepted that the gravitational effects
are too weak to affect -in an essential way- the properties of
soliton solutions of nonlinear field theories. 2. The resulting
equations are usually formidable such that the ordinary analytical
approaches become idle. More recently, however, the availability
of high speed computers and advanced numerical methods have
changed the case, and extensive numerical attempts have been made
in this direction (see e.g. 387N Term Project in \cite{5a}). It
has emerged from recent studies that the effects due to the
inclusion of gravity are not always negligible. Consider, for
example, the Einstein-Yang-Mills (EYM) system. It has been shown
that the EYM equations have both soliton and black hole solutions
(\cite{1}, \cite{2} and \cite{6}). This is in contrast to the fact
that vacuum Einstein and pure Yang-Mills equations do not have by
themselves soliton solutions. We can therefore conclude that
gravity may have dramatic effects on the existence or
non-existence of soliton solutions of nonlinear field equations.
Another interesting example is the discovery that black hole
solutions may have Skyrmion hair \cite{9}. It was previously
believed that stationary black holes can only have global charges
given by surface integrals at spatial infinity (the so-called
no-hair theorem).

In the ordinary O(3) model, spherically symmetric solutions have
an energy density which behave like $1/r^2$ at large distances
(\cite{7}). When formulated in a curved background, this model
leads to a spacetime which is  not asymptotically flat, and the
ADM mass is not well defined (\cite{5}). A nonlinear O(3) model
(thereafter referred to as the isovector model) was introduced in
(\cite{8}), which possesses spherical, soliton-like solutions with
a $1/r^4$ energy density  at large distances. Such a model, is
therefore expected to be well behaved in an asymptotically flat
background. In the present paper, we examine this model, and
discuss its self-gravitating solutions. These new solutions are
compared with those obtained previously in a flat spacetime.

The present manuscript is organized in the following way. In
section \ref{sec2}, we will review the isovector model of
\cite{8}. In section \ref{sec3}, flat-space solitons of the
isovector model and their resemblence to charged particles are
introduced. In section \ref{sec4}, the isovector model will be
reformulated in a curved background. The resulting differential
equations for a spherically symmetric ansatz will be introduced in
this section, together with the necessary boundary conditions.
These equations will be solved numerically, for several choices of
the coupling constant. We will compare the self gravitating
solutions with those obtained for a flat spacetime. Soliton masses
using the Tolman formalism will be discussed in section
\ref{sec5}, together with the behavior as a function of the model
parameter. Section \ref{sec6} will contain the summary and
conclusion.
\section{Isovector model \label{sec2}}
Consider an isovector field $\phi_a$ ($a=1,2,3$) with a $S^2$
vacuum at
\begin{equation}
\phi_a\phi_a=\phi_o^2.
\end{equation}
Each component $\phi_a$ is a pseudo-scalar under spacetime
transformations, and $\phi_o$ is a constant. A topological current
can be defined for such a field according to (\cite{8})
\begin{equation}
\label{topcur} J^\mu =\frac{1}{2\pi} \epsilon^{\mu \nu \alpha
\beta } \epsilon_{abc} \partial_\nu \phi_a
\partial_\alpha \phi_b \partial_\beta \phi_c.
\end{equation}
For the time being, spacetime is assumed to be the flat Minkowski
spacetime and $\mu ,\nu,... =0,1,2,3$ with $x^o= t$ ($c=1$ is
assumed throught this paper). $\epsilon^{\mu\nu\alpha\beta}$ and
$\epsilon_{abc}$ are the totally anti-symmetric tensor densities
in 4 and 3 dimensions, respectively. It can be easily shown that
the current (\ref{topcur}) is identically conserved ($\partial_\mu
J^\mu =0$), and the total charge is quantized
\begin{equation}
\label{topcharge} Q=\int J^o d^3x =\frac{1}{2\pi} \oint
\frac{dS_\phi}{dS_x} dS_x =ne,
\end{equation}
where $e\equiv 2\phi_o^3$. In this equation, $dS_x$ and $dS_\phi$
are area elements of $S^2$ surfaces in the $x$-space (as
$r\rightarrow \infty$) and $\phi$-space (as $\phi\rightarrow
\phi_o$), respectively. The current (\ref{topcur}) can identically
be written as the covariant divergence of an anti-symmetric,
second-rank tensor
\begin{equation}
\label{max1}
\partial_\mu F^{\mu \nu } =J^\nu,
\end{equation}
where
\begin{equation}
\label{feq} F^{\mu \nu} =\frac{1}{2\pi } \epsilon^{\mu \nu \alpha
\beta} \left[ \epsilon_{abc} \phi_a\partial_\alpha \phi_b
\partial_\beta \phi_c +\partial_\beta {\mathcal C}_\alpha \right],
\end{equation}
in which ${\mathcal B}_\alpha $ is an auxiliary vector field. The
dual field $\ ^*F$ with the tensorial components
\begin{equation}
\ ^*F^{\mu \nu} =\frac{1}{2} \epsilon^{\mu \nu \alpha \beta }
F_{\alpha \beta } =\frac{1}{4\pi}(2\epsilon_{abc} \phi_a
\partial^\mu \phi_b
\partial^\nu \phi_c +\partial^\nu {\mathcal C}^\mu -\partial^\mu
{\mathcal C}^\nu) ,
\end{equation}
satisfies the equation
\begin{equation}
\label{max2}
\partial_\mu \ ^*F^{\mu \nu } = 0,
\end{equation}
provided that the vector field ${\mathcal C}^\mu$ is a solution of
the following wave equation
\begin{equation}
\Box {\mathcal C}^\mu  -\partial^\mu (\partial_\alpha {\mathcal
C}^\alpha ) =2\epsilon_{abc} \partial _\nu ( \phi_a
\partial^\nu \phi_b \partial^\mu \phi_c  ) .
\end{equation}
It can be easily shown that the right hand side of this
equation defines another conserved current
\begin{equation}
K^\mu = 2\epsilon_{abc}\partial_\alpha ( \phi_a \partial^\mu \phi_b
\partial^\alpha \phi_c ),
\end{equation}
with
\begin{equation}
\partial_\mu K^\mu =0.
\end{equation}
Using the language of differential forms, (\ref{feq}) can be
written in the following form
\begin{equation}
F=G+H,
\end{equation}
where the components of the 2-forms $G$ and $H$ are given by
\begin{equation}
G_{\mu \nu} =\frac{1}{2\pi } \epsilon^{\alpha\beta}_{\ \ \ \mu\nu}
\epsilon_{abc} \phi_a \partial_\alpha \phi_b \partial_\beta \phi_c
,
\end{equation}
and
\begin{equation}
H_{\mu\nu} =\frac{1}{2\pi } \epsilon_{\mu\nu}^{\ \ \ \alpha\beta}
\partial_\beta \mathcal{B}_\alpha .
\end{equation}
We now have
\begin{equation}
\label{df0}
dF=0,
\end{equation}
and
\begin{equation}
d^*H =0.
\end{equation}
The 2-form $F$ is therefore Hodge-decomposable, and cohomologous
with $G$ (i.e. they belong to the same cohomology class, since
they differ only by an exact form). The resemblance of equations
(\ref{max1}) and (\ref{max2}) to the Maxwell's equations and the
capability of this model to provide non-singular solutions
behaving like charged particles were discussed in \cite{8}. In the
next section, we will only outline the main results valid in a
flat spacetime.
\section{Flat space solitons\label{sec3}}
The requirement of having non-singular, finite energy and stable
solitons, severely restrict the possible choices of the lagrangian
density of the isovector field. Let us follow \cite{8}, and adopt
the following lagrangian density which satisfies the above
requirements:
\begin{equation}
\label{isolag}
{\cal L}=\lambda \left( \partial_\mu \phi_a \partial^\mu \phi_a \right)^2
-b_o (1-\frac{\phi}{\phi_o})^2,
\end{equation}
with $\lambda <0$, and $b_o$ real constants. The potential $V(\phi
)=b_o (1-\phi/\phi_o)^2$ satisfies the following conditions
\begin{equation}
V(\phi_o)=0, \ \ \ (\frac{\partial V}{\partial \phi})_{\phi_o}=0, \ \ \
{\rm and}\ \ \ (\frac{\partial^2 V}{\partial \phi^2})_{\phi_o}>0,
\end{equation}
which leads to the spontaneous breaking of the (global) $SO(3)$ symmetry
of the system. The dynamical equation for the isovector field is easily obtained,
using the variational principle $\delta \int {\cal L}d^4x=0$, which leads to
\begin{equation}
\label{isoeq}
\partial_\mu \partial_\nu \phi_b
\partial^\mu \phi_a
\partial^\nu \phi_b +\partial_\nu \phi_b \partial_\mu
\partial_\nu \phi_b \partial^\mu \phi_a
+\partial_\nu \phi_b \partial^\nu \phi_b \Box \phi_a
=-\frac{1}{4\lambda}\frac{\partial V}{\partial \phi_a}.
\end{equation}
Similar to the ansatz used in the Skyrme model (\cite{10}), we
start with the so-called hedgehog ansatz
\begin{equation}
\label{hedge}
\phi_a=\phi (r)\frac{x^a}{r},
\end{equation}
where $x^a$, $a=1,2,3$ represent the Euclidean coordinates $x$, $y$,
and $z$, respectively. This ansatz immediately leads to
\begin{equation}
K^\mu=0, \ \ {\cal B}^\mu=0, \ \  \vec{B}=0,
\end{equation}
and
\begin{equation}
\vec{E}=\frac{2\phi^3}{r^2}\hat{r}=\frac{e}{r^2}\left(
\frac{\phi}{\phi_o}\right)^3\hat{r},
\end{equation}
where $\hat{r}$ is the unit vector in the radial direction, and
$e$ is the elementary topological charge defined  in
(\ref{topcharge}). Note that $F^{oi}=E_i$ and
$F^{ij}=-\epsilon_{ijk}B_k$. The charge density corresponding to
this ansatz is easily obtained to be
\begin{equation}
\rho=J^o=\frac{3}{2\pi}\frac{\phi^2}{r^2} \frac{d \phi}{d r},
\end{equation}
which leads to
\begin{equation}
Q=\int \rho 4\pi r^2 dr =2\phi_o^3 \equiv e,
\end{equation}
showing that the ansatz bears unit topological charge. Note that
in deriving this result, we have used the boundary conditions
\begin{equation}
\phi_a(r=0)=0,
\end{equation}
and
\begin{equation}
\phi \rightarrow \phi_o , \ \ \ \rm{as} \ \ \ r\rightarrow \infty
.
\end{equation}
It can be shown that the following asymptotic solutions are valid:
\begin{equation}
\phi(r) \simeq \phi_o\left[ \alpha_o \frac{r}{r_o}
-\frac{1}{10\alpha_1^2} \left( \frac{r}{r_o}\right)^2+...\right] ,
\end{equation}
close to the center of the soliton ($r\rightarrow 0$), where
$\alpha_1$ is a dimensionless constant, and
\begin{equation}
\phi(r)\simeq \phi_o
\end{equation}
and
\begin{equation}
\vec{E}\simeq \frac{e}{r^2}\hat{r},
\end{equation}
far from the soliton (i.e. $r\rightarrow \infty$).

It can be seen that the total soliton energy
\begin{equation}
M=\int T^o_o d^3 x =\int_o^\infty \left[  -\lambda \left(
(\frac{\partial \phi}{\partial r})^4 +4\frac{\phi^4}{r^4}
+4\frac{\phi^2}{r^2}(\frac{d\phi}{dr})^2 \right) +V(\phi)\right]
4\pi r^2 dr=5.21\phi_o^3\left[ b_o(-4\lambda)^3\right]^{1/4}.
\end{equation}
Using the scale transformation $r\rightarrow \alpha r$ (while
keeping $\phi$ unchanged), it can be shown that $M(\alpha )$ has a
minimum at $\alpha=1$, which is a signature of the stability of
the soliton under radial perturbations.
\section{Self-gravitating isovector solitons\label{sec4}}
By self-gravitating isovector solitons, we mean static solutions
of the coupled isovector-gravitational equations which are
everywhere regular and represent localized lumps of energy. By
numerically integrating the coupled nonlinear equations, we will
show that such solutions do arise depending on the value of the
model parameters. Based on or results, we will also criticize the
widely expressed view that gravity has only a minute effect on the
structure and properties of extended solitons.

Let us start with the action\begin{equation}
\label{action}
{\cal A}=\int \left( -\frac{R}{16\pi G} +{\cal L}_M \right) \sqrt{-g}
d^4x,
\end{equation}
in which $G$ is the gravitational constant, $R$ is the
curvature scalar, ${\cal L}_M$ is the lagrangian
density of the matter source, and
$g$ is the determinant of the metric tensor. As the source of gravity, we
consider the isovector field (\ref{isolag}).
By varying the action (\ref{action}) with respect to $g_{\mu\nu}$ and $\phi_a$,
we obtain the corresponding field equations:
\begin{equation}
\label{einseq}
R_{\mu\nu}=8\pi G\lambda \left[
    4(\partial^\beta \phi_b\partial_\beta \phi_b )
         (\partial_\mu \phi_a\partial_\nu \phi_a )
         -g_{\mu\nu}(\partial^\beta \phi_b\partial_\beta \phi_b )^2
         -g_{\mu\nu}\frac{V(\phi)}{\lambda} \right] ,
\end{equation}
and
\begin{equation}
(\partial^\beta \phi_b\partial_\beta \phi_b )\Box \phi_a
+g^{\mu\nu}\partial_\mu\phi_a\partial_\nu
(\partial^\beta \phi_b\partial_\beta \phi_b )
=-\frac{1}{4\lambda}\frac{\phi_a}{\phi}\frac{\partial V(\phi)}{\partial \phi}.
\end{equation}
By contracting equation (\ref{einseq}), we obtain the following equation
for the curvature scalar:\begin{equation}
R=-32\pi GV(\phi).
\end{equation}
This equation is useful, since it expresses a simple relation
between the curvature scalar and the self-interaction potential of
the isovector field, and provides a means to check some of the
calculations. We employ the coordinates $x^\mu =(t,r,\theta,
\phi)$, and the general, spherically symmetric, static metric
$g_{\mu\nu}=\rm{diag} (-B(r), A(r), r^2, r^2\sin \theta)$. For the
hedgehog ansatz (\ref{hedge}), the independent field equations
become
\begin{equation}
\label{met1} \frac{B'}{B}+\frac{A'}{A}=-64\pi G \lambda \left(
\frac{r\phi'^4}{2A}+\frac{\phi^2\phi'^2}{r}\right)^2,
\end{equation}
\begin{equation}
\label{met2} \frac{B'}{B}-\frac{A'}{A}=-64\pi G\lambda \left(
\frac{r\phi'^4}{4A}-\frac{A\phi^4}{r^3}+\frac{rAV(\phi)}{4\lambda}
\right) +\frac{2A}{r}-\frac{2}{r},
\end{equation}
and
\begin{equation}
\label{isograv}
\frac{3\phi'^2\phi''}{A^2}+\frac{2\phi^2\phi''}{r^2A}-\frac{3A'\phi'^3}{
    2A^3}+\frac{B'\phi'^3}{2A^2B}+\frac{2\phi'^3}{rA^2}
    +\frac{2\phi \phi'^2}{r^2A}-\frac{A'\phi^2\phi'}{r^2A^2}
    +\frac{B'\phi^2\phi'}{r^2AB}-\frac{4\phi^3}{r^4}
    =-\frac{1}{4\lambda}\frac{\partial V}{\partial \phi}.
\end{equation}
Note that equation (\ref{isograv}) reduces to the flat-space
equation (\ref{isoeq}), by putting $A=B=1$. Equations (\ref{met1})
and (\ref{met2}), however, become inconsistent for obvious reasons
(the spacetime cannot be flat in the presence of matter sources).

Let us introduce the following three length scales
\begin{equation}
r_o\equiv \left( \frac{-4\lambda}{b_o}\right)^{1/4}\phi_o,
\end{equation}
\begin{equation}
r_{g1}\equiv \frac{1}{4\sqrt{\pi Gb_o}},
\end{equation}
and
\begin{equation}
r_{g2}\equiv \sqrt{-\pi G\lambda}\phi_o^2,
\end{equation}
which appear naturally in equations (\ref{met1}) to
(\ref{isograv}). It can be seen that $r_o$ is proportional to the
geometric mean of $r_{g1}$ and $r_{g2}$:
\begin{equation}
r_o=\sqrt{\frac{r_{g1}r_{g2}}{2}}.
\end{equation}

 Emergence of new length scales is similar to what happens in
non-abelian gauge fields (\cite{12}), and leads to the appearance
of new branches of spherical, static solitons. Having additional
length scales, one expects departures from the Einstein-Maxwell
solutions and the appearance of more subtle features. Using these
two length scales, equations (\ref{met1}) to (\ref{isograv}) can
be made dimensionless and suitable for numerical integration;
\begin{equation}
\label{metdim1}
\frac{B'}{B}+\frac{A'}{A}=\frac{1}{\epsilon}\left( \frac{xu'^4}{2A}+\frac{
    u^2u'^2}{x}\right)^2,
\end{equation}
\begin{equation}
\label{metdim2}
\frac{B'}{B}-\frac{A'}{A}=\frac{1}{\epsilon}\left( \frac{xu'^4}{4A}-\frac{
    Au^4}{x^3}-xAW(u)\right) +\frac{2A}{x}-\frac{2}{x},
\end{equation}
and
\begin{equation}
\label{isogravdim}
\frac{3u'^2u''}{A^2}+\frac{2u^2u''}{x^2A}-\frac{3A'u'^3}{
    2A^3}+\frac{B'u'^3}{2A^2B}+\frac{2u'^3}{xA^2}
    +\frac{2u u'^2}{x^2A}-\frac{A'u^2u'}{x^2A^2}
    +\frac{B'u^2u'}{x^2AB}-\frac{4u^3}{x^4}
    =\frac{\partial W(u)}{\partial u}.
\end{equation}
In these equations, $u=\phi /\phi_o$, $x=r/r_o$, and
$\epsilon\equiv \alpha^2$ which is the  dimensionless parameter of
the system:
\begin{equation}
\label{epseq} \epsilon =\left(
\frac{r_{g1}}{r_o}\right)^2=\frac{1}{32\pi G\phi_o^2\sqrt{-\lambda
b_o}},
\end{equation}
and
\begin{equation}
\label{weq} W(u)=(1-u)^2.
\end{equation}

The asymptotic behavior of the non-singular solutions can be found by
using the following series ansatze:
\begin{equation}
u(x)=\sum_{i=0}^\infty c_ix^i, \ \ A(x)=\sum_{i=0}^\infty a_ix^i, \ \
B(x)=\sum_{i=0}^\infty b_ix^i,
\end{equation}
for $x\rightarrow 0$, and
\begin{equation}
u(x)=1+\sum_{i=1}^\infty \frac{c'_i}{x^i}, \ \
A(x)=1+\sum_{i=1}^\infty \frac{a'_i}{x^i}, \ \
B(x)=1+\sum_{i=1}^\infty \frac{b'_i}{x^i},
\end{equation}
for $x\rightarrow \infty$. The unknown coefficients are calculated
by inserting these ansatze into equations (\ref{metdim1}) to
(\ref{isogravdim}) and balancing the terms of the same order in
$x$. Table \ref{coeff} shows the leading coefficients for the
above asymptotics.

\begin{table}
\begin{tabular}{|l||l|}
\hline $r\rightarrow 0$ & $r\rightarrow \infty$ \\
\hline\hline
$a_o= 1$  & $a'_2=-\frac{1}{2\epsilon}$ \\
$a_2=\frac{1}{24}\frac{9\omega^4-4}{\epsilon}$ &
$a'_4= \frac{1}{4\epsilon^2}$\\
$a_3=\frac{1}{10}\frac{\omega}{\epsilon}$&
$a'_6=-\frac{1}{40}\frac{5-16\epsilon^2}{\epsilon^3}$ \\
$a_4=\frac{1}{72000}\frac{12555\omega^{10}-12240\omega^6-2880\omega^4\epsilon
+2000\omega^2-936\epsilon}{\epsilon^2\omega^2}$ &
$a'_8=\frac{5-32\epsilon^2}{80}$ \\ \hline
$b_o= 1$ & $b'_2=\frac{1}{2\epsilon}$ \\
$b_2=\frac{1}{24}\frac{9\omega^4+4}{\epsilon}$&
$b'_6=-\frac{2}{5\epsilon}$\\
$b_3=-\frac{13}{30}\frac{\omega}{\epsilon}$&
$b'_{10}=-\frac{68}{45\epsilon}$\\
$b_4=\frac{1}{8000}\frac{1035\omega^{10}+220\omega^6+1440\omega^4\epsilon
+268\epsilon}{\epsilon^2\omega^2}$&
$b'_{12}=-\frac{96}{55\epsilon^2}$\\
\hline
$c_1=\omega$ & $c'_4=-2$ \\
$c_2=-\frac{1}{10\omega^2}$&
$c'_8=-28$\\
$c_3=\frac{1}{750}\frac{45\omega^{10}-35\omega^6
+30\omega^4\epsilon-9\epsilon}{\omega^5\epsilon}$&
$c'_{10}=-\frac{28}{\epsilon}$\\
$c_4=\frac{1}{1350000}\frac{35505\omega^{10}+6260\omega^6+10920\omega^4\epsilon
-3456\epsilon}{\epsilon\omega^8}$& $c'_{12}=-1648$\\ \hline
\end{tabular}
\caption{Leading, non-vanishing coefficients of the asymptotic
solutions.} \label{coeff}
\end{table}

\begin{figure}[t]
 \epsfxsize=10cm
  \centerline{\epsffile{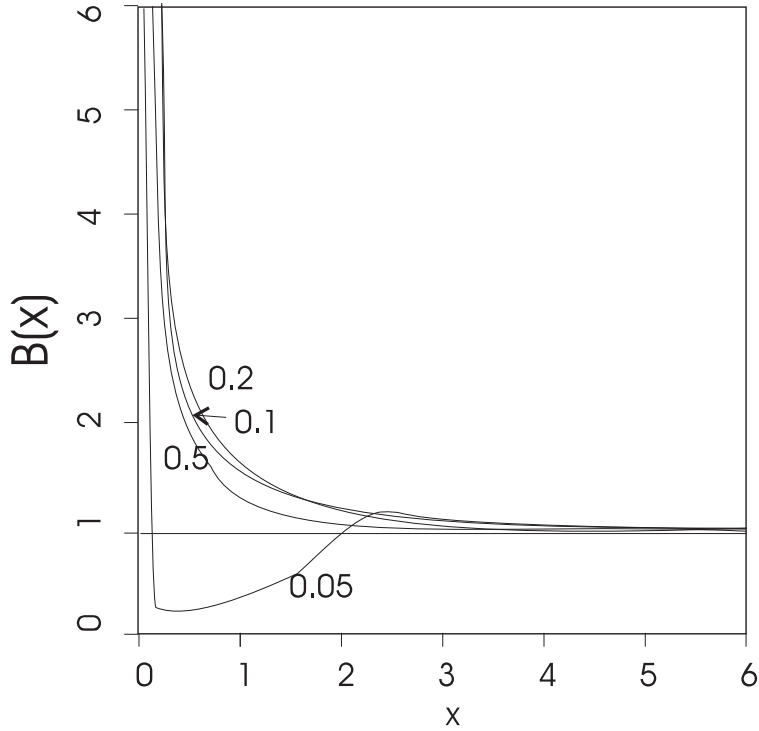}}
  \caption{ Variations of the isovector field amplitude for
several values of the parameter $\epsilon$. Downward: flat space,
$\epsilon$=0.2,0.1, and 0.05.}
  \label{fig1}
\end{figure}

In order to solve the coupled nonlinear DEs numerically, we used
the Gerald's shooting method, which is based on two guesses for
the initial slope of the unknown functions. Using these initial
guesses, the equations are integrated via the Runge-Kutta-Fehlberg
method, reaching the end point of the independent variable. A
better guess for the initial slope is then found by comparing the
end point values with the boundary conditions, and interpolating
for the initial slope (\cite{12}). The boundary conditions for
asymptotically flat spacetime read
\begin{equation}
u\rightarrow 0 \ \ {\rm as} \ \ x\rightarrow 0,
\end{equation}
and
\begin{equation}
u\rightarrow 1,\ \ A\rightarrow 1,\ \ {\rm and} \ \ B\rightarrow
1, \ \ {\rm as} \ \ x\rightarrow \infty .
\end{equation}
The procedure described above is iterated until the correct
boundary values are reached with a reasonable accuracy.

\begin{figure}[t]
 \epsfxsize=10cm
  \centerline{\epsffile{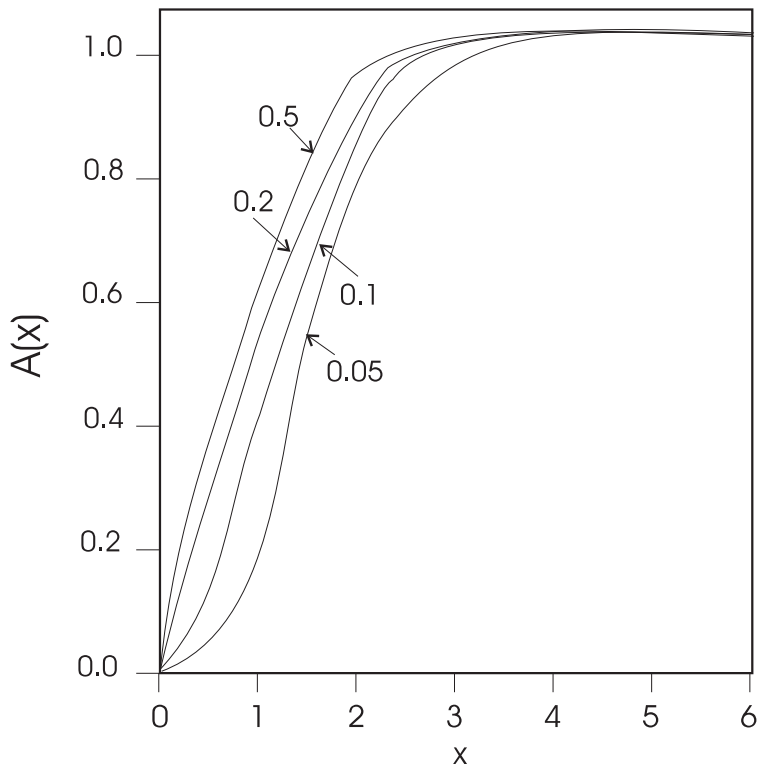}}
  \caption{ Variations of the metric coefficient A for
several values of the parameter $\epsilon$ (=0.5,0.2,0.1, and 0.05
downward).}
  \label{fig2}
\end{figure}

\begin{figure}[t]
 \epsfxsize=10cm
 \centerline{\epsffile{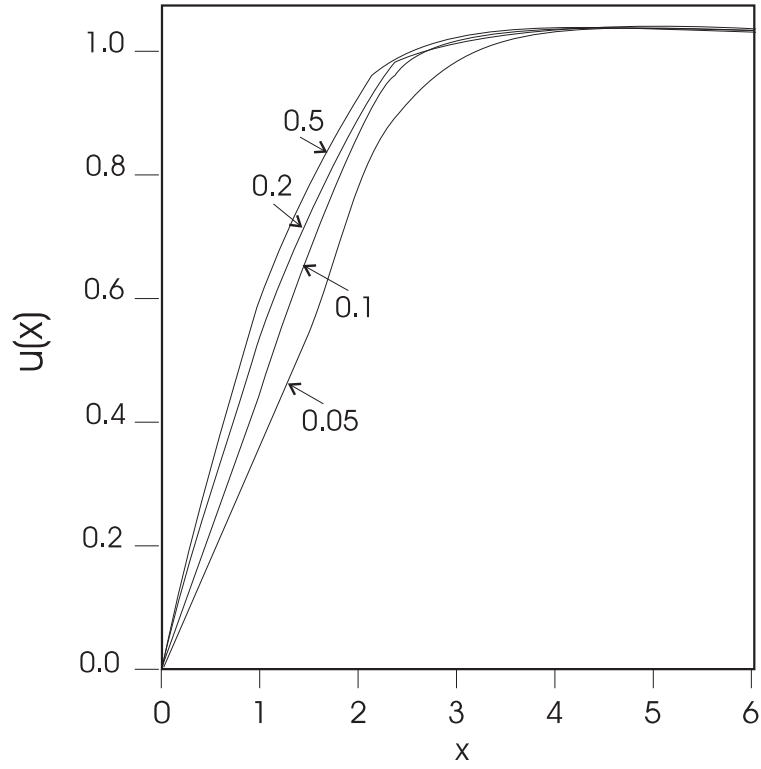}}
  \caption{ Variations of the metric coefficient B for
several values of the parameter $\epsilon$ (=0.2,0.1,0.5, and 0.05
downward).}
  \label{fig3}
\end{figure}

In order to test the method, flat space solutions were first
computed and compared to the solutions obtained via energy
minimization algorithms (\cite{8}). Figure \ref{fig1} shows
numerical variations of the  $u(x)$ function for several values of
the parameter $\epsilon$. The corresponding results for the metric
coefficients $A(x)$ and $B(x)$ are shown in Figures \ref{fig2} and
\ref{fig3}, respectively.

It is seen that for $\epsilon$ of the order of unity, the
self-gravitating solutions differ only slightly from the
flat-space solution. The difference vanishes completely as
$\epsilon\rightarrow \infty$. As $\epsilon$ becomes much smaller
than 1, significant differences with the flat-space solution
emerge. For example, the metric signature changes at some $r$ for
sufficiently small $\epsilon$.

\section{Soliton masses\label{sec5}}
Although there are still controversies about an exact definition
for the total mass of a self-gravitating system (\cite{13},
\cite{14},\cite{15}), we adopt the Tolman formalism for computing
the total mass of the gravitating solitons;
\begin{equation}
M_T =\int I_T \sqrt{-g}d^3x,
\end{equation}
where
\begin{equation}
I_T=T^o_o-T^1_1-T^2_2-T^3_3,
\end{equation}
and
\begin{equation}
T^{\mu\nu}=\frac{2}{\sqrt{-g}}\left[ \frac{\partial
(\sqrt{-g}L_M)}{\partial
g_{\mu\nu}}-\frac{d}{dx^\sigma}\frac{\partial
(\sqrt{-g}L_m)}{\partial g_{\mu\nu,\sigma}} \right] .
\end{equation}

\begin{figure}[t]
 \epsfxsize=10cm
  \centerline{\epsffile{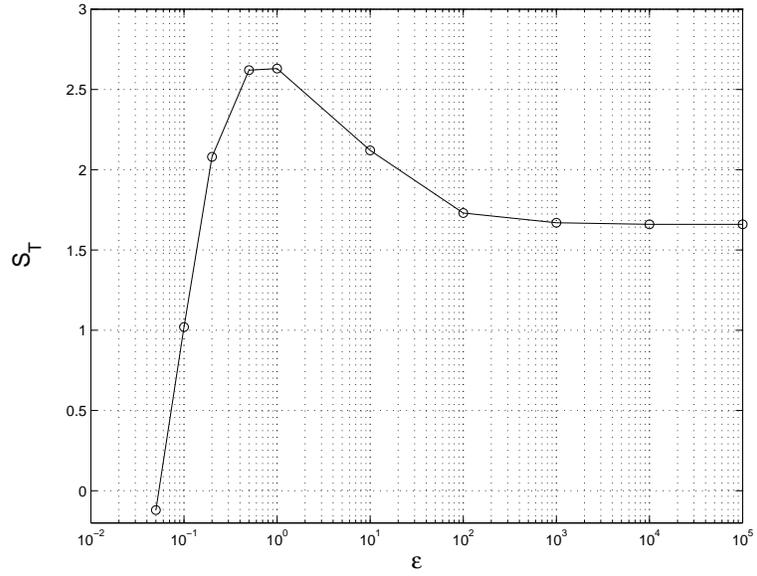}}
  \caption{ Variations of the integral $S_T$ as a function of the dimensionless parameter
  $\epsilon$.}
  \label{massfig}
\end{figure}

In most cases (including the model we are discussing here), the
second term is absent since the matter lagrangian does not contain
derivatives of the metric tensor. For the isovector model, we have
\begin{equation}
I_T=4\lambda \left(
\frac{\phi'^2}{A}+\frac{2\phi^2}{r^2}\right)\left( \frac{2r\phi
'\phi ''}{A}-\frac{rA'\phi'^2}{A}+\frac{4\phi \phi
'}{r}-\frac{4\phi^2}{r^2}\right) -2r\phi'\frac{\partial
V}{\partial \phi}-\left( \frac{rA'}{A}+\frac{rB'}{B}+6\right)
\left[ -\lambda \left( \frac{\phi '
}{A}+\frac{2\phi^2}{r^2}\right)^2 +V(\phi)\right].
\end{equation}
Using the transformations (\ref{epseq}) and (\ref{weq}), the total
mass can be written in terms of a dimensionless integral:
\begin{equation}
\label{tolmass} M_T=\pi \phi_o^3 [b_o(-4\lambda)^3 ]^{1/4}S_T,
\end{equation}
where
\[
S_T=-4\int_o^\infty ( \left(
\frac{u'^2}{A}+\frac{2u^2}{A}+\frac{2u^2}{x^2}\right) \left(
\frac{2xu'u''}{A}-\frac{xA'u'^2}{A^2}-\frac{u'^2}{A}-\frac{4uu'}{x}
-\frac{4u^2}{x^2}\right) +
\]
\begin{equation}
\left( 6+\frac{xA'}{A}+\frac{xB'}{B}\right)\left(
W(u)+\frac{1}{4}\left( \frac{u'^2}{A}+\frac{2u^2}{r^2}\right)^2
\right)+2xu'\frac{\partial W(u)}{\partial u})  \sqrt{AB}x^2 dx.
\end{equation}
and $\eta =-4\lambda$. The dimensionless integral $S_T$ were
computed numerically for several values of the parameter
$\epsilon$. The results are shown in Figure \ref{massfig}.

It is seen that (1) $S_T\rightarrow 1.66$ as $\epsilon \rightarrow
\infty$, leading to the asymptotic (flat space) mass, (2) There is
a maximum mass around $\epsilon \simeq 1$, and (3) The total mass
decreases as $\epsilon \rightarrow 0$, with $M_T\simeq 0$ at
$\epsilon \simeq 0.055$. It is also interesting to note that in
the asymptotic series solutions summarized in Table \ref{coeff},
the coefficients $a'_o$ and $b'_0$ vanish, which imply vanishing
total mass of the soliton as deduced from the asymptotic form of
the metric \cite{17}.

\section{Summary and conclusion\label{sec6}}

We extended the isovector model to incorporate the effects of
gravity. The resulting equations were integrated numerically for
spherically symmetric ansatz, using the Gerald's shooting method.
It was found that for large values of the dimensionless parameter
of the system the effect of gravity is negligible. For small
values of $\epsilon$, however, gravity has a considerable effect
on the qualitative and quantitative behavior of the solutions.
Such dramatic changes in the behavior of the spherical solutions
in the presence of gravity were also reported in the framework of
EYM equations (\cite{1}, \cite{2}, and \cite{6}). Gravitating
solitons of the isovector model in an asymptotically flat
background bear quantized topological charges, exactly similar to
the flat-space solitons. The quantization is due to a $\pi^2$
homotopy between the boundary of the curved space ($S^2$ at $r
\rightarrow \infty$), and the vacuum $S^2$ of the isovector field
($\phi_a\phi_a =\phi_o^2$). This is in analogy with the
quantization of the magnetic pole intensity in the t'Hooft
Polyakov monopoles (\cite{16}). Solutions presented in this paper
(Figures \ref{fig1} to \ref{fig3}) do not exhibit horizons. Using
the well-known result from general relativity (\cite{17}), $\tau_2
/\tau_1=\sqrt{B(r_2)/B(r_1)}$, where $\tau_1$ and $\tau_2$ are the
emitted (at $r_1$) and detected (at $r_2$) periods of photons,
event horizons correspond to $B(r_1)=0$, which are not fulfilled
by the present solutions. However, the appearance of horizons and
signature changes is not ruled out  and will be addressed
elsewhere. In particular, it should be interesting to know whether
isovector black holes have hair. As pointed out in \cite{11},
solitons can make bound states with ordinary black holes to form
hairy black holes. In such a case, the total ADM mass of the hairy
black hole is the sum of the mass of the bare black hole, the mass
of the soliton, and the gravitational binding energy between the
two (\cite{11}).

\acknowledgements N. Riazi acknowledges the support of Shiraz
University (research grant 80-SC-1424-C152) and IPM.


\end{document}